# Multi-source Domain Adaptation Using Gradient Reversal Layer for Mitotic Cell Detection

Satoshi Kondo

Muroran Institute of Technology, Hokkaido 050-8585, Japan

**Abstract.** This is a write-up of our method submitted to Mitosis Domain Generalization (MIDOG 2021) Challenge held in MICCAI2021 conference.

## 1    Introduction

**Motivation** Mitosis detection is a key component of tumor prognostication for various tumors, including breast cancer. Scanning microscopy slides with different scanners leads to a significant visual difference, resulting in a domain shift. This domain shift prevents most deep learning models from generalizing to other scanners, leading to strongly reduced performance.

**Scope** Detect mitotic figures (cells undergoing cell division) from histopathology images (object detection) [1]. Images scanned by 4 different scanners, 3 out of which are labeled, are provided. In total the set consists of 200 cases of breast cancer. Evaluation will be done on four scanners (two new, two are part of the training set) with the F1 score as main metric.

## 2    Proposed Method

Our method is two-step approach. The first step is extraction of candidate regions of mitosis. In the second step, we classify the candidate regions to mitosis and non-mitosis. We explain the details of each step in the following.

### 2.1    Extraction of candidate regions

The input RGB images are first transformed into blue ratio (BR) images. The blue ratio image, which accentuates the nuclear dye, is computed as the ratio of the blue channel and the sum of the other two channels [2]. Candidate mitotic regions are extracted by binary thresholding of the BR image. The regions are cropped as rectangle patches and these patches are the candidate for mitosis.

## 2.2 Classification of candidate regions

The candidate regions are classified to mitosis or non-mitosis. We have histopathological images scanned by four different scanners, three out of which are labeled. By using the labeled images, we train a deep neural network model which has two classification tasks. The first task is mitosis / non-mitosis classification, which is a binary classification, and the second one is scanner classification, which is three-class classification. We use ResNet [3] as the base model. We remove the final fully connected layer in ResNet and append two branches at the end of the network. The first branch is for mitosis / non-mitosis classification and three fully connected layers. The second branch is for scanner classification and Gradient Reversal Layer [4] followed by another three fully connected layers.

In the training phase using labeled source domain data, we use all patches extracted from the images of three scanners. Eighty percent of images for each scanner are used for training data and the rest images are used for validation data. We use cross entropy loss for each classification and the final loss is the summation of two losses.

In the domain adaptation phase, we use the images of three scanners, which are labeled and treated as the source domain data, and one image from a target domain which has the target image for mitosis detection.

In general setting of unsupervised domain adaptation, all the source domain data are used in domain adaptation phase, but it's not feasible since we must keep all source domain images. Therefore, we select patches in source domain and use those selected patches in the domain adaptation phase. We use the classifier trained using source domain images to select the patches. We select the patches which have high confidence (high probability) as mitosis or non-mitosis when the patches are classified by the classifier.

In the domain adaptation phase, we extract candidate regions from the image from the target domain at first. We have no labels on mitosis / non-mitosis for these patches, but we have labels on scanner. We treat these patches as coming from the fourth scanner (fourth class in the scanner classifier). We use cross entropy loss for mitosis / non-mitosis classification and scanner classification for the source domain data, and cross entropy loss for scanner classification for the target domain data. The final loss is the summation of those losses.

After the domain adaptation phase finishes, the patches in the target domain are classified to mitosis or non-mitotic regions.

## 3 Experimental Conditions

In the extraction of the candidate regions, the threshold is set to the mean value plus 3 times of the standard deviation of the BR image. Regions that are smaller than 2 pixels, its width is longer than 50 pixels or its height is longer than 50 pixels are eliminated. The size of a patch is 64 x 64 pixels.

For example, we selected about 873k patches from 50 images of scanner 1. In these patches, 672 patches were mitotic regions.

We use 18-layer ResNet as the base network. In the training of the classifier using the source domain data, the learning rate is 1.0e-5, which are optimized with Optuna library [5], the batch size is 128, the number of epochs is 30. The optimizer is Adam [6] and the learning rate changes with cosine annealing.

We select 10k patches from each scanner in the source domain for the domain adaptation phase. Ten out of 10k patches are mitotic regions.

In the domain adaptation phase, the learning rate is 1.7e-6, which are optimized with Optuna library using source domain data, the batch size is 128, the number of epochs is 5. The optimizer is Adam and the learning rate does not change.

## References

1. https://midog2021.grand-challenge.org/
2. Chang, H., Loss, L. A., Parvin, B.: Nuclear segmentation in H&E sections via multi-reference graph cut (MRGC). In International symposium biomedical imaging. (2012).
3. He, K., Zhang, X., Ren, S., Sun, J.: Deep residual learning for image recognition. In Proceedings of the IEEE conference on computer vision and pattern recognition, pp. 770-778 (2016).
4. Ganin, Y., Ustinova, E., Ajakan, H., Germain, P., Larochelle, H., Laviolette, F., Marchand, M., Lempitsky, V.: Domain-adversarial training of neural networks. The journal of machine learning research, 17(1), 2096-2030 (2016).
5. Akiba, T., Sano, S., Yanase, T., Ohta, T., Koyama, M.: Optuna: A next-generation hyperparameter optimization framework. In Proceedings of the 25th ACM SIGKDD international conference on knowledge discovery & data mining, pp. 2623-2631 (2019).
6. Kingma, D. P., Ba, J.: Adam: A method for stochastic optimization. In: 3rd International Conference on Learning Representations, ICLR 2015, San Diego, CA, USA, May 7-9, 2015, Conference Track Proceedings (2015).